\documentclass{bioinfo}

\usepackage[square,numbers,compress]{natbib}
\usepackage{booktabs}
\usepackage{multirow}
\usepackage{amsmath}
\usepackage{graphics}
\usepackage{pifont}  
\usepackage{tikz}
\usepackage{subcaption}
\usepackage{float}
\usepackage{array}
\usepackage{url}
\usepackage{tcolorbox}
\usepackage{xcolor,etoolbox}

\copyrightyear{2020} \pubyear{2020}

\access{Advance Access Publication Date: Day Month Year}
\appnotes{Manuscript Category}

\begin{document}
\firstpage{1}

\subtitle{Subject Section}

\title[Survey for disease gene prediction]{Recent Advances in Network-based Methods for Disease Gene Prediction}
\author[Sezin Ata \textit{et~al}.]{Sezin Kircali Ata\,$^{\text{\sfb 1}}$, Min Wu\,$^{\text{\sfb 2}}$, Yuan Fang\,$^{\text{\sfb 3}}$, Le Ou-Yang\,$^{\text{\sfb 4}}$, Chee Keong Kwoh\,$^{\text{\sfb 1}}$ and Xiao-Li Li\,$^{\text{\sfb 2,}*}$}
\address{$^{\text{\sf 1}}$School of Computer Science and Engineering, Nanyang Technological University, 639798, Singapore \\
$^{\text{\sf 2}}$Institute for Infocomm Research, 138632,
Singapore \\
$^{\text{\sf 3}}$School of Information Systems, Singapore Management University, 188065,
Singapore and \\
$^{\text{\sf 4}}$College of Information Engineering, Shenzhen University, Shenzhen 518060, China}

\corresp{$^\ast$To whom correspondence should be addressed.}

\history{Received on XXXXX; revised on XXXXX; accepted on XXXXX}

\editor{Associate Editor: XXXXXXX}

\abstract{Disease-gene association through Genome-wide association study (GWAS) is an arduous task for researchers. Investigating single nucleotide polymorphisms (SNPs) that correlate with specific diseases needs statistical analysis of associations. Considering the huge number of possible mutations, in addition to its high cost, another important drawback of GWAS analysis is the large number of  false-positives. Thus, researchers search for more evidence to cross-check their results through different sources. To provide the researchers with alternative and complementary low-cost disease-gene association evidence, computational approaches come into play. Since molecular networks are able to capture complex interplay among molecules in diseases, they become one of the most extensively used data for disease-gene association prediction. In this survey, we aim to provide a comprehensive and up-to-date review of network-based methods for disease gene prediction. We also conduct an empirical analysis on 14 state-of-the-art methods. To summarize, we first elucidate the task definition for disease gene prediction. Secondly, we categorize existing network-based efforts into network diffusion methods, traditional machine learning methods with handcrafted graph features and graph representation learning methods. Thirdly, an empirical analysis is conducted to evaluate the performance of the selected methods across seven diseases. We also provide distinguishing findings about the discussed methods based on our empirical analysis. Finally, we highlight potential research directions for future studies on disease gene prediction.
}

\maketitle

\section*{Introduction}

Genetic diseases are mostly caused by gene mutations, although recent studies reveal that epigenetic factors can also play a role~\citep{Zenk212}. Among the existing methods, linkage analysis and genome-wide association studies (GWAS) are the most fundamental approaches in disease gene prediction, as they can provide predictive biomarkers through the genetic variation studies among humans, known as single nucleotide polymorphisms (SNPs). Nevertheless, statistical analysis and biological validation of the biomarkers are costly and time consuming, as a large amount of false positives need to be analyzed further \citep{snpnewstat2018}. Additionally, these techniques are simply based on direct genotype-phenotype associations. However, biological molecules perform their functions in corresponding pathways in a collaborative fashion. Projecting and characterizing their specific roles and collaborations onto a wired network/graph structure can reveal more useful knowledge and provide more systematical aspects. Furthermore, in a network-based environment, the disease causing factors, such as genetic mutations, epigenetic factors and pathogens, can be tracked more efficiently by chasing network perturbations, i.e., edge or node removals, in the molecular networks \citep{perturbations}. Therefore, molecular networks are efficient and effective data representations, which are able to model complex interplay among the molecules through a wider viewpoint and track the potential disruptions on the biological pathways due to the disease-causing factors. As such, they have been extensively used by the computational approaches to complement and enrich existing linkage analysis and GWAS studies.

Currently, there are several molecular networks available to describe relationships between genes, such as protein-protein interaction (PPI) networks, gene regulatory networks, gene co-expression and metabolic interaction (MI) networks. Among these networks, PPI networks are the most extensively leveraged for disease-gene association prediction. One key reason is that proteins perform a vast array of critical functions to sustain an organism's well-being. Some of these functions include biochemical reactions,  metabolic reaction catalysing, DNA replication, transmission of signals between cells, and maintaining structure for cells and tissues. More specifically, proteins collaborate and interact with each other to perform biological functions, leading to many protein interactions, which can be integrated and modelled as a graph/network data structure. Given a protein interaction, a corresponding gene mutation from one of the two proteins could make existing interaction impossible, and thus loses certain important biological functions and causes diseases.  
Another reason is, several studies using a PPI network embody the assumption that the position of a protein is not random. Proteins associated with a common set of biological properties tend to have common topological properties in the network such as node degree and centrality \citep{nettop1,Ideker2008}. Therefore, PPI networks can be employed in revealing protein-disease associations through the useful network-based features for proteins. 
However, existing PPI networks are incomplete, i.e., only a fraction of real protein interactions are detected through  high-throughput experiments. Likewise they are noisy due to biased experimental evidence towards much studied disease genes \citep{Menche2015}. Thus, an integrative approach covering various aspects of proteins such as GWAS, gene-expression, gene ontology, and other domain knowledge is both important and necessary. 

In this survey, we focus on reviewing the computational methods leveraging network/graph data for disease gene prediction. First, we introduce the problem definition (i.e., node classification and link prediction) for disease gene prediction with different types of graph inputs. Second, we classify the network-based methods into three categories  and provide a brief introduction for these methods. Third, we select representative methods from each category and conduct a comprehensive empirical study on them. The three categories are listed in the following.
\begin{itemize} 
    \item \textbf{Network diffusion methods}. The diffusion methods employ random walk techniques for influence propagation in different networks (e.g., PPI networks or phenotype-gene networks) for disease gene prediction.
    \item \textbf{Machine learning methods with handcrafted graph features}. Various features for diseases and genes are first extracted from input graph data, and then fed into traditional machine learning models (e.g., Random Forest) for predicting disease-gene associations.
    \item \textbf{Graph representation learning methods}. Instead of using handcrafted features for disease gene prediction, graph representation learning methods automatically learn the latent features or embeddings for diseases and genes by matrix factorization, graph embedding and graph neural network techniques.
\end{itemize}

Former surveys on network-based methods 
\citep{perturbations,Ideker2008,barabasi2011network}, 
which were published about ten years ago, present pioneering endeavours on network  analysis and bring to light the importance of network data for disease gene prediction. Nowadays, we witness the usefulness of incorporating network data in several areas ranging from drug discovery to disease gene identification through novel network-based algorithms. In this survey, we aim to bring together up-to-date methods for disease gene prediction and provide a wider perspective to this important problem. In addition, there are two recent surveys which review the recent network embedding efforts on biomedical networks \citep{SuEmbSurvey2018, yue2020graph}. In \citep{SuEmbSurvey2018}, the authors present an overview of the existing network embedding methods and their applications in biomedical data science, e.g., pharmaceutical data analysis, multi-omics data analysis and clinical data analysis. However, they do not conduct any empirical evaluation for the introduced methods. In  \citep{yue2020graph}, the authors introduce the recent graph embedding methods and their applications in biomedical networks. In particular, they further select 11 graph embedding methods and perform a systematic evaluation on 5 different tasks, i.e., drug-disease association prediction, drug-drug interaction prediction, PPI prediction, protein function prediction and medical term semantic type classification. Yet, they do not touch the task on disease gene prediction and its various state-of-the-art approaches. 
In this survey, we aim to conduct an empirical analysis on disease gene prediction task using different methods from the above three categories. Besides, we further apply and evaluate recent graph embedding methods including heterogeneous network embedding and multi-view network embedding for disease gene prediction.

The rest of this survey is organized as follows. In Section `Problem Definition for Disease Gene Prediction', we cast the task of disease gene prediction as an instance of node classification or link prediction based on different graph inputs. Then, we introduce various network-based disease-gene prediction methods in details in Section `Network-based Methods for Disease Gene Prediction'. Next, in Section `Empirical Comparison', we perform a comprehensive empirical evaluation for 14 representative methods. Finally, in Section `Future Perspectives', we discuss potential future research directions in disease-gene association prediction problem and we conclude this paper in Section `Conclusion'.


\section*{Problem Definition for Disease Gene Prediction} \label{sec:Tasks}

\begin{figure*} [t]
			\centering
	\begin{subfigure}[b]{.4\columnwidth}
	\includegraphics[scale=0.2]{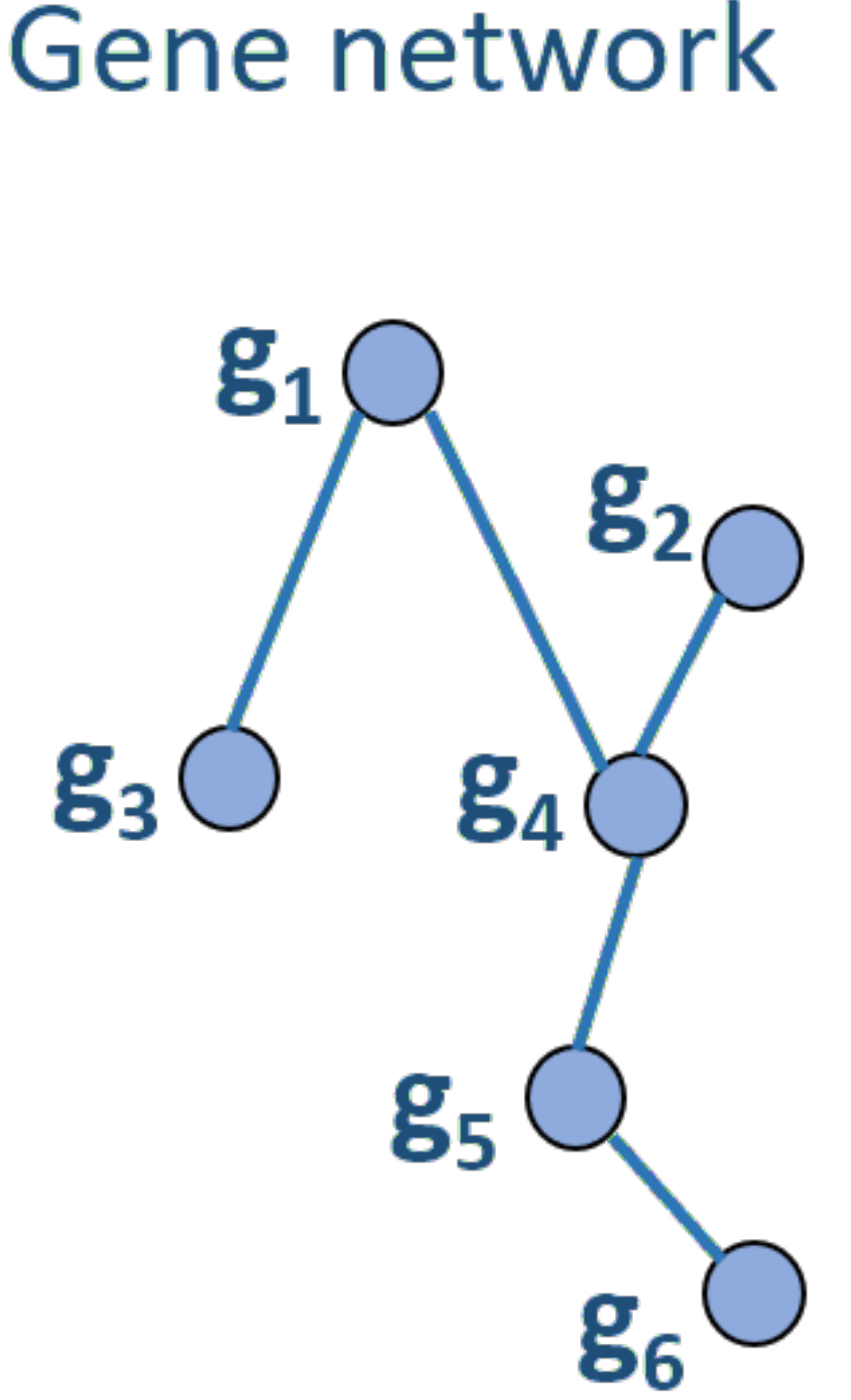}
	\caption{A homogeneous graph.} 
	\end{subfigure}
	\begin{subfigure}[b]{.5\columnwidth}
		\centering
	\includegraphics[scale=0.35]{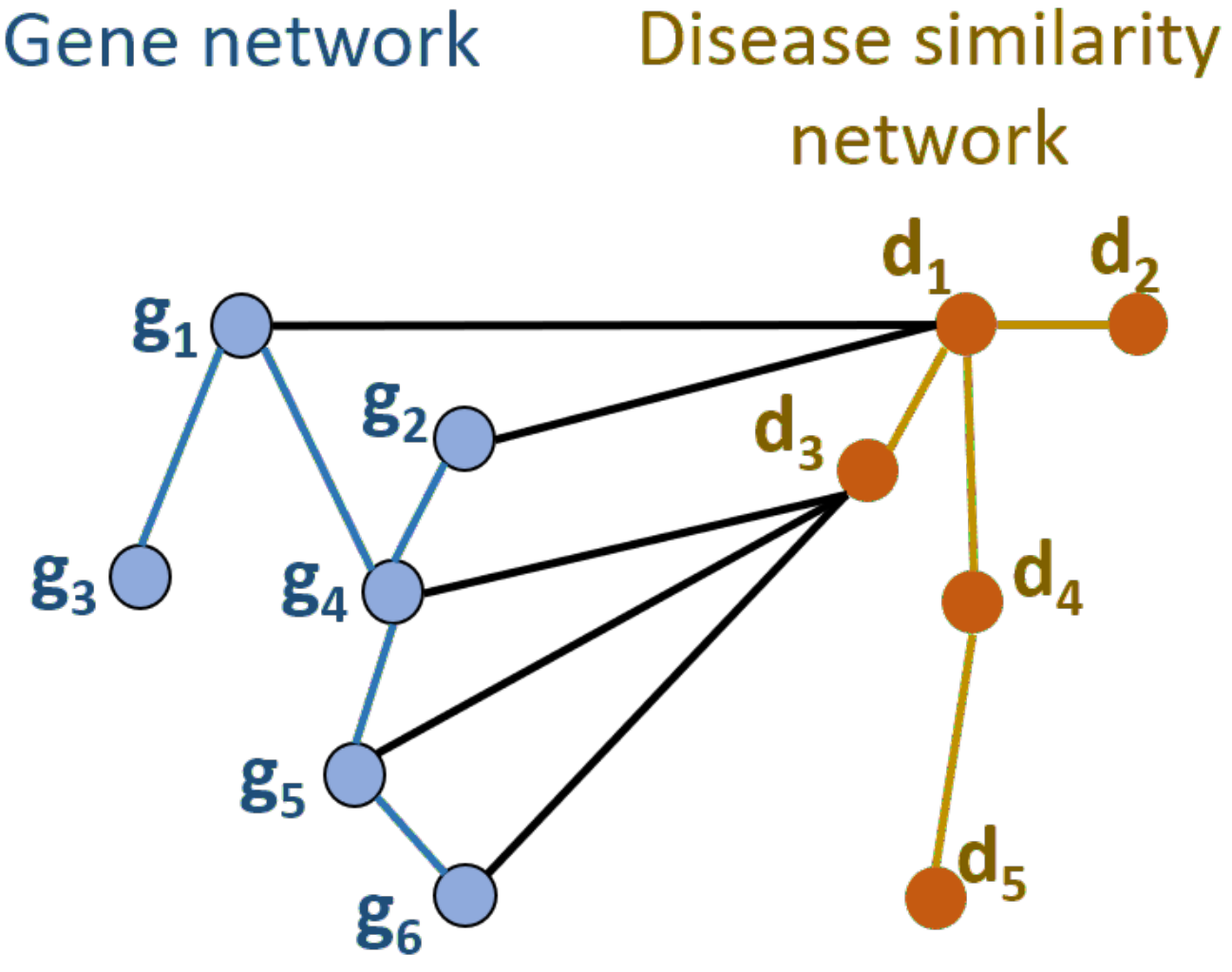}	\caption{A heterogeneous graph.} 
	\end{subfigure}
	\begin{subfigure}[b]{\columnwidth}
	\centering
	\includegraphics[scale=0.45]{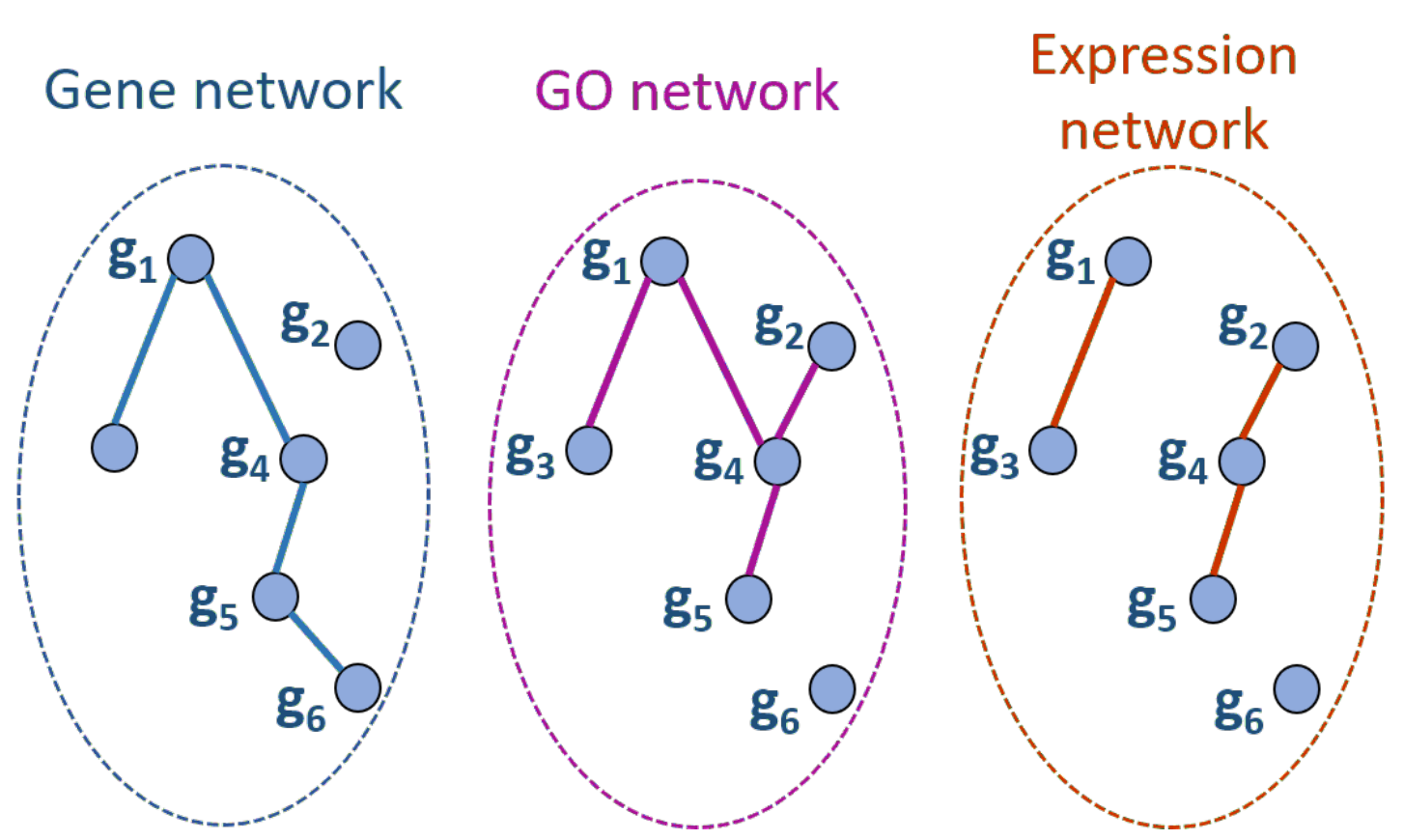}	
	\caption{A multi-view graph with three views.} 
	\end{subfigure}
    \caption{Different types of graph inputs for network-based methods: (a) homogeneous graph, (b) heterogeneous graph and (c) multi-view graph.}
	\label{fig:inputs}
\end{figure*}

Network-based disease gene prediction leverages graph/network data as inputs to predict disease-causing genes. In particular, different types of graphs have been exploited for this purpose, including homogeneous graphs, heterogeneous graphs and multi-view/multiplex graphs. A homogeneous graph refers to a graph with a single type of nodes and a single type of edges, while a heterogeneous graph contains different types of nodes and edges. In addition, a multi-view or multiplex graph is a collection of graphs with the same set of nodes and different types of edges (e.g., edges from different views). Figure \ref{fig:inputs} shows the examples of different types of graph inputs. PPI network in Figure \ref{fig:inputs}(a) is a homogeneous graph, while phenotype-gene network in Figure \ref{fig:inputs}(b) is a heterogeneous graph. Figure \ref{fig:inputs}(c) shows a multi-view graph for proteins, containing three views from the perspectives of PPI, GO similarity and gene expression.

Based on the above graph inputs, network-based disease gene prediction can be treated as a node classification or link prediction task. Node classification aims to infer the disease label of the unlabelled genes by utilizing the known disease genes, whereas link prediction aims to predict disease causing genes by utilizing gene-disease associations. Next, we give a formal definition for these two tasks from the perspective of disease gene prediction.

\subsection*{Node Classification}

Figure~\ref{fig:tasks}(a) shows the node classification tasks, which is to predict the label of the genes of which disease associations are unknown, given known labels on some genes/nodes.

More formally, assume that we have a homogeneous graph $G = (V, E)$ where $V$ is the set of nodes/genes and $E$ shows relationships between nodes. A subset of genes $V_{labeled} \subset V$ represents known disease-causing genes, while $V_{unknown} = V \setminus V_{labeled}$ represents the set of genes of which disease associations are unknown. The formulation of node classification on $G$ is to predict the labels for the nodes in $V_{unknown}$. This node classification task in a heterogeneous graph or multi-view graphs can be similarly defined.

\subsection*{Link Prediction}

The link prediction is a common task to reveal relationships between the objects especially in recommendation systems and social network analysis. Given a network $G$, denoted as $G=(V,E)$ and the nodes $v_i, v_j \in V$, the task of link prediction is to provide a measure of proximity/similarity between the nodes $v_i$ and $v_j $. 

In particular, we consider a heterogeneous graph $G=(U,V,E)$ for gene-disease association prediction. The vertices are grouped into two sets $U$ and $V$, representing the set of genes and the set of diseases, respectively. $E$ includes the edges in $U$, the edges in $V$ and the edges between $U$ and $V$ (i.e., known disease-gene associations). The goal of gene-disease association prediction is to predict unknown links between $U$ and $V$. 

\begin{figure}[!htb]
	\begin{subfigure}[b]{.53\columnwidth}
	\includegraphics[scale=0.30]{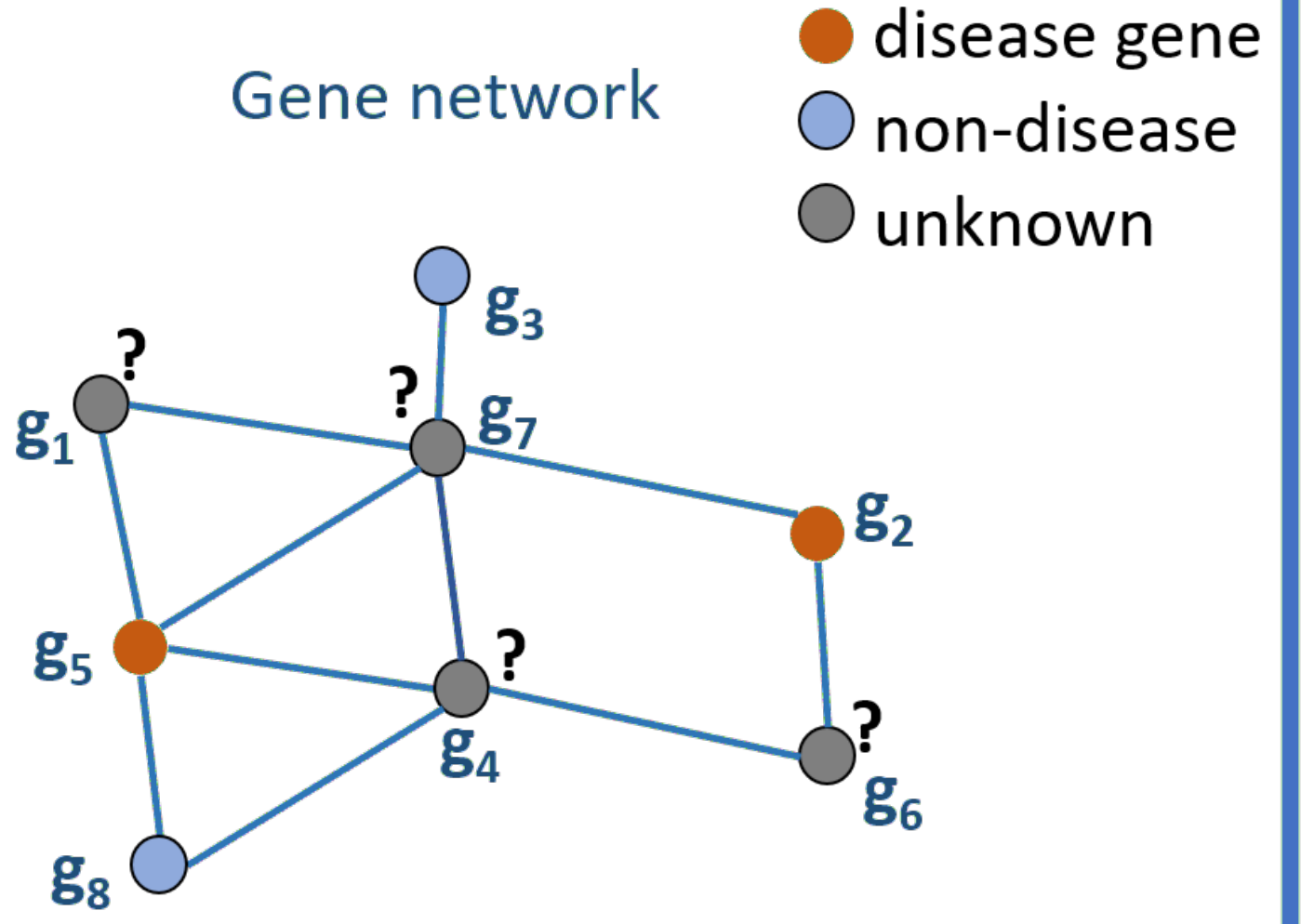}
	\caption{Node classification.} 
	\end{subfigure}
	\begin{subfigure}[b]{.39\columnwidth}
	\includegraphics[scale=0.27]{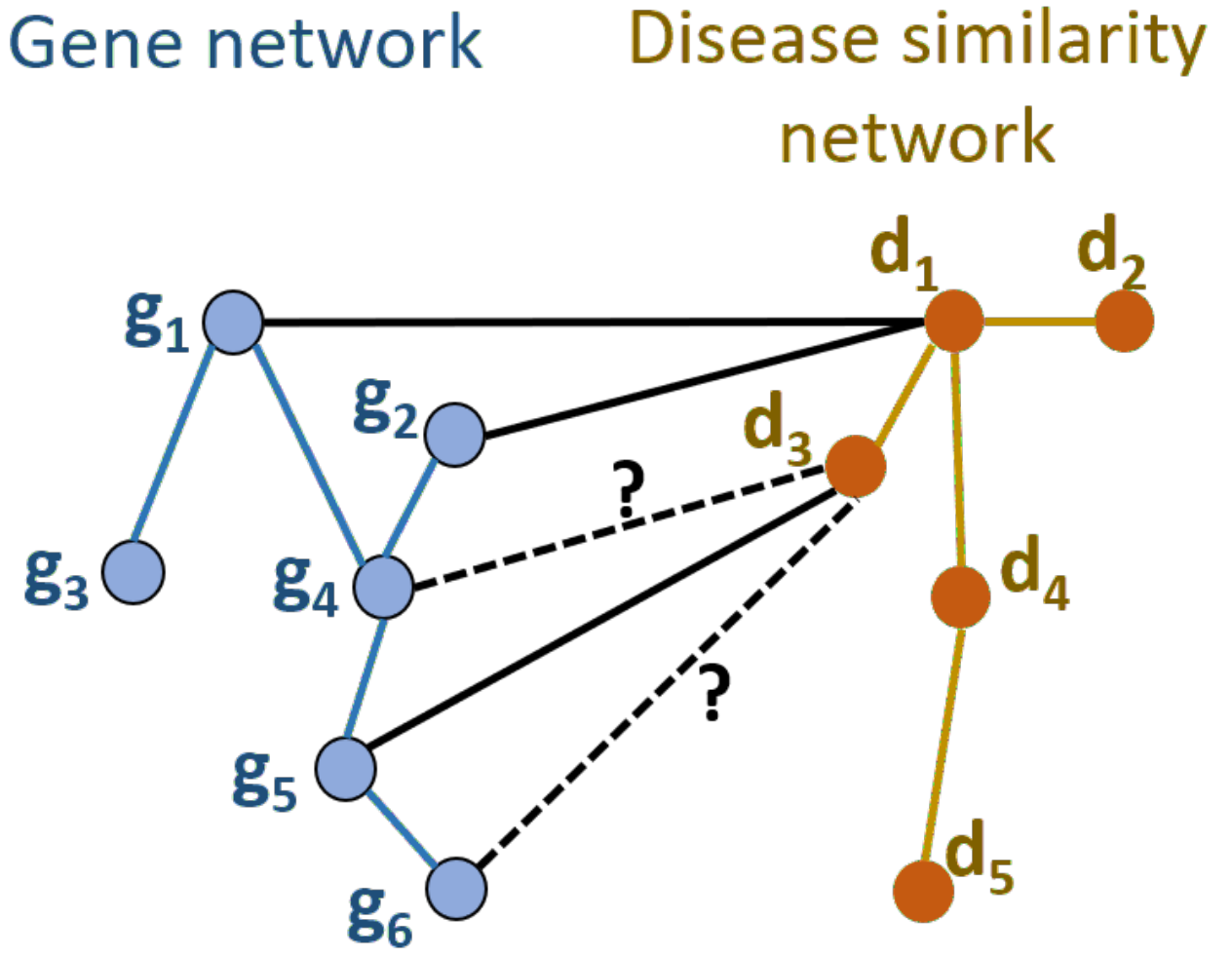}	\caption{Link prediction.} 
	\end{subfigure}
    \caption{Tasks in disease gene prediction.}
	\label{fig:tasks}
\end{figure}

Figure~\ref{fig:tasks}(b) shows the disease-gene association prediction in a heterogeneous disease-gene network. The edges in black show the existing associations between diseases and genes. Meanwhile, the dashed line represents a potential association between the pair of $g_4$ and $d_3$, which is to be predicted by computational methods.

\section*{Network-based Methods for Disease Gene Prediction} \label{sec:model_review}

\begin{figure*}[tbp]
 	\centering
 	\includegraphics[scale=1.1]{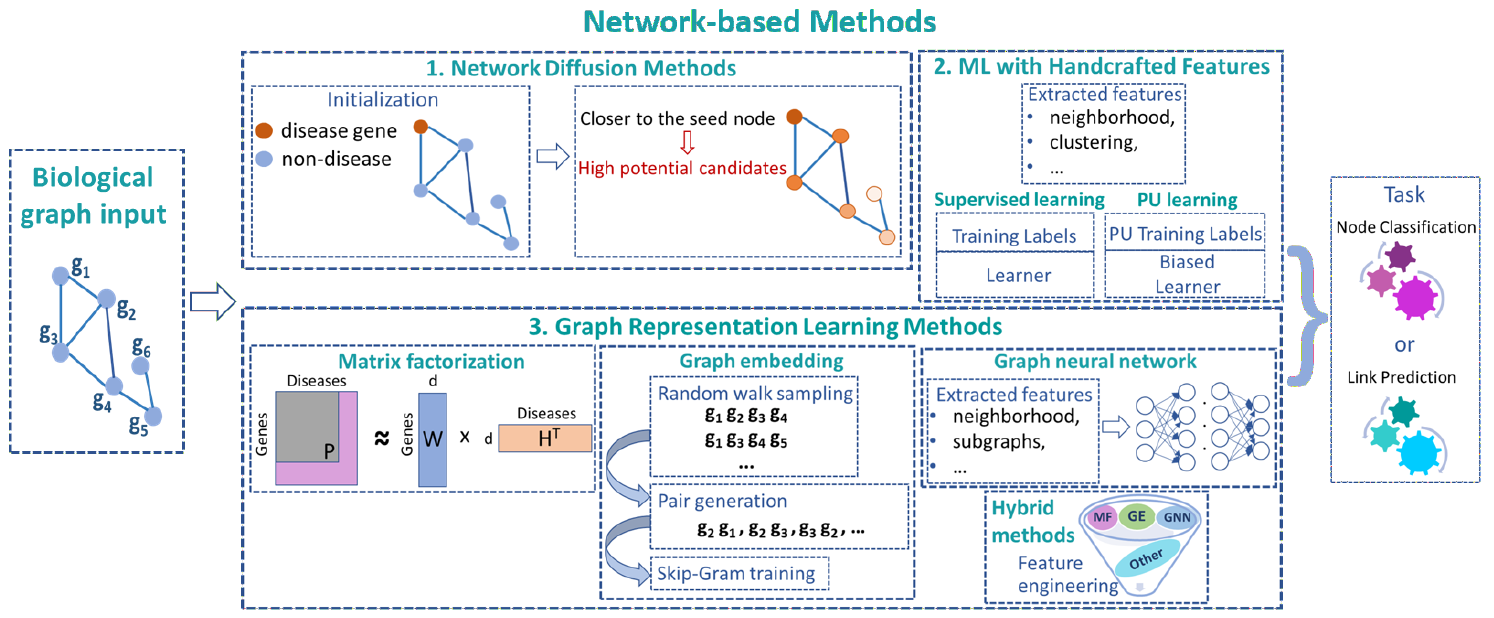}
 	\caption{Pipelines for the categories in network-based studies for disease gene prediction.}
 	\label{fig:Categories}
 \end{figure*}

In this section, we present a comprehensive review of network-based methods for disease gene prediction, namely diffusion-based methods, traditional feature-based methods and graph representation learning methods as illustrated in Figure~\ref{fig:Categories}. 


\subsection*{Network Diffusion Methods}

Diffusion-based methods working on biological networks are mostly adopted from pioneer graph-based semi-supervised algorithms such as \citep{semiPioneer1,RWREsas}. In disease gene prediction problem, they are widely exploited in the analysis of the biological pathways especially the pathways that are closest to the known disease genes. Starting from known disease genes, they diffuse along the biological network through random walks. Next, we introduce diffusion methods by walking in different networks. 

\subsubsection*{Random Walk in PPI Network}

Random walk with restart (RWR) is the most extensively used algorithm especially in prioritization of disease genes. Differently than a basic random walk, it is able to jump back to any seed node with probability $ r $ at each iteration. Formally, it is defined as follows:

\begin{equation}\label{eq:RWR}
p^{t+1}=(1-r)Wp^{t}+rp^{0},
\end{equation}

where $ W $ is the column-wise normalized adjacency matrix of the network. In particular, $ p^{t} $ is the probability vector of being at node $ i $ at time step $ t $ in its $ i $-th entry and $ p^{0} $ 
is the initial probability vector holding the probabilities of being at known disease nodes (seeds) \citep{RWR}. Initially, each known disease protein has a uniformly distributed probability as the sum of the probabilities equal to 1. Total number of iterations is determined by the condition which satisfies the L1 norm difference between $ p^{t} $
and $ p^{t+1} $ smaller than a pre-defined threshold (e.g., $ 10^{-6} $).

PRINCE \citep{PRINCE} adopts RWR to a weighted PPI network through a weight function and also utilizes the disease similarities as prior probabilities. On an unweighted PPI network, PRINCE basically performs random walk with restart (Eq.~\ref{eq:RWR}) with adjusted prior probabilities based on the disease similarity scores. VAVIEN \citep{Vavien2011rwrexample} is proposed to measure the topological similarity of proteins by formulating their interactions with Pearson correlation coefficient as a topological profile. To compute this profile, they employ the random walk proximity as a feature between seed and candidate proteins, so that the proteins with similar interactions will have a similar topological profile. Then, they prioritize the candidate disease genes based on the topological profile scores. In \citep{Orient2013(rwrexample)}, the authors introduce a method called ORIENT, which prioritizes the candidate disease genes with RWR. They perform RWR on the adjacency matrix where the weights of interactions close to the known disease genes are properly reinforced by computing the shortest path distance to the disease genes. In \citep{diffusionStrong}, the authors propose a method called DP-LCC to construct diffusion profiles for the disease genes and the candidate genes separately based on their RWR on the PPI and the phenotype similarity network. Candidate genes are prioritized according to their diffusion profile similarities with the query disease. NPDE \citep{2015PrinceEsentialPro} utilizes non-disease essential proteins by formulating a dual flow network propagation method to prioritize candidate disease genes. It uses the assumption that if non-disease essential proteins exist as neighbors of a candidate protein, the protein is unlikely to cause a disease. Eventually it formulates this assumption as the negative flow in the prior information of the network propagation and the positive flow is allocated for the disease proteins.

\begin{table*}[b]
\centering
\caption{Various network diffusion methods for disease gene prediction.}
\begin{tabular}{llp{11cm}}  \hline
Categories & Methods & Method descriptions \\\hline
\multirow{5}{*}{\shortstack{RW in PPI network \\(Node classification)}} 
& PRINCE \citep{PRINCE} & Network propagation on a PPI network.\\
& VAVIEN \citep{Vavien2011rwrexample} & Random walk and correlation on a PPI network.\\
& ORIENT \citep{Orient2013(rwrexample)} & Neighbor-favoring RWR on a PPI network.\\
& DP-LCC \citep{diffusionStrong} & RWR-based diffusion profiles based on PPI and phenotype similarity network.  \\
& NPDE \cite{2015PrinceEsentialPro} & Network propagation with dual flow on a PPI network.\\
\hline
\multirow{3}{*}{\shortstack{RW in heterogeneous network \\(Node classification)}}  
& BioGraph \citep{BioGraph2011(RWRdiffusionBased)} & Stochastic RW on an integrated network containing 21 publicly available curated databases. \\

& CrossRank \citep{TissuePropogate2016Koyuturk} & Optimization based on network propagation on a heterogeneous network with various aspects.
\\
& SLN-SRW \citep{SupervisedRwr2017} & Supervised RW to learn edge weights of the integrated network and RWR for prediction. \\\hline
\multirow{5}{*}{\shortstack{RW in heterogeneous network \\(Link prediction)}}
    & RWRH ~\citep{RWRH} & RWR on a heterogeneous phenotype-gene network.\\
    & RWPCN ~\citep{supPropagation} & RWR on a heterogeneous network with phenotypes, genes and protein complexes.\\
    & RWRH-Malaria \citep{diffusionExpMalaria} & RWR on cross-species PPI networks for human and parasite.\\
    & BiRW \citep{Bi-RandomWalk2015} & RWR on a heterogeneous network with a phenotype, gene and the phenotype-gene associations.\\
    & RWRMH ~\citep{RWRMH2018} & RWR on a multiplex heterogeneous network of protein interactions and disease networks.\\
\hline\end{tabular}
	\label{tab:sum_diffusion}%
\end{table*}

\subsubsection*{Random Walk in Heterogeneous Network}

For diffusion methods in heterogeneous networks, we further divide them into two categories with respect to their task as node classification or link prediction.

We start with node classification methods. BioGraph, \citep{BioGraph2011(RWRdiffusionBased)}, is a data integration and data mining platform. It utilizes stochastic model of random walks with restarts for a given candidate gene prioritization query as incorporating multiple data sources such as disease, pathway and GO annotation. In \citep{TissuePropogate2016Koyuturk}, two methods, CrossRank and CrossRankStar, formulate disease gene prioritization problem as optimization problems based on network propagation. They model two types of networks, network of networks (NoN) for CrossRank and networks of star networks (NoSN) model for CrossRankStar and both models incorporate tissue specific molecular networks. In SLN-SRW \citep{SupervisedRwr2017}, the authors first aim to integrate biomedical data from heterogeneous sources including multiple ontologies and databases. For this propose, a simplified Laplacian normalization based supervised random walk algorithm is employed to learn the edge weights of the integrated network. RWR is then performed on this integrated network for disease gene prediction.

Link prediction methods aim to predict disease-gene associations in heterogeneous phenotype-gene network, where PPI network and phenotype network are connected through known phenotype-gene relationships. RWRH \citep{RWRH} aims to prioritize proteins based on their relevance to disease proteins. Basically, RWRH extends RWR algorithm to the heterogeneous phenotype-gene network through inter and intra transitions between PPI network and phenotype network. It is performed for a given query disease and corresponding disease genes are considered as seed nodes. In \citep{supPropagation}, a RWRH-based method called RWPCN is proposed to predict and prioritize disease genes on an integrated network comprising human protein complexes, protein interaction network, and phenotype similarity network. The basic difference from RWRH is that RWPCN operate random walks in an additional protein-complex networks. Likewise, a RWRH-based approach to predict malaria-associated genes on an integrated network by integrating human-human, parasite-parasite and human-parasite protein interactions in \citep{diffusionExpMalaria}. BiRW~\citep{Bi-RandomWalk2015}, aims to capture circular bigraph patterns based on the assumption that the relation among phenotype-gene associations can be characterized by these patterns. For this purpose, they employ a bi-random walk algorithm and aim to reconstruct the missing associations globally through the gene-disease association prediction. In a recent study \citep{RWRMH2018}, the authors propose two extensions of RWRH, which are RWRM and RWRMH. Former method performs RWR on a multiplex network (i.e., all nodes are same type) composed of three layers of networks containing PPI, co-expression and pathway associations of proteins. Latter method incorporates a disease-disease network based on phenotype similarities, and gene-disease bipartite associations in addition to the aforementioned multiplex network as multiplex-heterogeneous network so that the random walker can jump to a network containing different sets of edges and nodes. Overall, a summary of different diffusion-based methods is presented in Table~\ref{tab:sum_diffusion}. 


\subsection*{Machine Learning Methods with Handcrafted Features}

\subsubsection*{Supervised machine learning methods}

\begin{table*}[b]
\centering
\caption{Machine learning methods with hand-crafted graph features for disease gene prediction.}
\begin{tabular}{llp{11cm}}  \hline
Categories & Methods & Method descriptions \\\hline
\multirow{7}{*}{Supervised learning}
& DERanking \citep{MLGenePriortz2010} & 4 ranking strategies based on whether a gene is surrounded by highly differentially expressed genes.\\
& BRIDGE \citep{2013PriortzGenesRgressn} & A regression model with lasso penalty to weight 5 genomic data sources.\\
& IMRF \citep{supDataIntegration}  & An improved MRF method on multiple networks, e.g., PPI, pathways, protein complexes, etc.\\
& Metagraph+ \citep{ata2017disease} & Using metagraph features extracted from a heterogeneous network with PPI and gene keywords.\\
& dgMDL \citep{LuoMDL2019} & Multi-modal deep belief nets on PPI network and gene network based on GO similarities. \\
& DiGI~\citep{node_kernel2020} & A regularized linear SVM on the features extracted from a novel decomposition kernel.\\
& NetWAS \citep{Greene2015}  & Integrated analysis through regularized Bayesian integration of tissue-specific networks with SVMs.\\

\hline
\multirow{5}{*}{PU learning}  
& ProDiGe \citep{Prodige} & Biased SVM using features derived from multiple data sources \citep{de2007kernel} including PPI.\\
& PUDI \citep{PUDI} &  Multi-level classification with biased SVM on features extracted from PPI, protein domain and GO. \\
& CATAPULT \citep{Catapult} & Biased SVM with features derived from walks in a heterogeneous phenotype-gene network.\\
& EPU \citep{yang2014ensemble} & Ensemble PU learning using various features derived from PPI and GO similarity networks.  \\
& PEGPUL \citep{jowkar2016perceptron} & A perceptron ensemble of graph-based PU learning from 3 base classifiers SVM, KNN and CART. \\

\hline\end{tabular}
	\label{tab:handcrafted-feature}%
\end{table*}

Supervised machine learning methods employ features and/or kernels to integrate various biological concepts, thus have been extensively studied in disease gene prediction. These methods work on the adjacency matrix of the network data. They extract features for genes/proteins based on various graph related measures such as shortest path distance, diffusion kernels, neighborhood with a disease protein, common neighbours, metapaths, metagraphs, etc. In \citep{MLGenePriortz2010}, DERanking is proposed to benchmark four different strategies and prioritize candidate genes according to network analysis of their differential expression. The reason of incorporating differential expression as prior knowledge is based on the assumption that the strong disease candidates tend to be surrounded by differentially expressed neighbors. Three of the four benchmarking strategies are distinct random walk-based strategies from exponential diffusion kernel approach and one of them is direct neighborhood analysis. These strategies are performed on four networks comprising functional or psychical interactions of proteins for disease gene prediction. In \citep{2013PriortzGenesRgressn}, a method called BRIDGE is introduced for prioritization of disease genes by integrating various gene aspects including PPI, protein sequence data, gene expression data, KEGG database and GO, through a weighting scheme. This scheme is attained through a multiple linear regression model with lasso penalty, which determines the phenotypic similarity between two diseases based on the functional similarities between their associating genes. Accordingly, the model is capable of identifying genes associated with the diseases whose genetic bases are completely unknown. In \citep{supDataIntegration}, multiple aspects of proteins including known gene-disease associations, protein complexes,
PPIs, pathways and gene expression profiles are integrated by Markov random field (MRF) and Bayesian analysis for disease gene prioritization. Initial prior probability of the MRF is set by Gibbs sampling. Metagraph representations \citep{ata2017disease} are proposed for disease gene prediction, which utilize both PPI network and biological annotations called \textit{keywords} through heterogeneous subgraphs. By counting the occurrences of proteins within a specific subgraph type in terms of connectivity patterns, feature vectors of proteins are constructed, so that the proteins, which are functionally similar but located far away in a PPI network still have a chance to have similar representations in case they co-occur within the same subgraph type. 
In~\citep{LuoMDL2019}, the authors employ a multimodal deep belief net named dgMDL to predict disease–gene associations for all known diseases instead of predicting associated genes for a specific disease. This could alleviate the risk of overfitting due to the small number of positives and a large number of features in disease gene prediction problem. They first train two multi-modal DBN, one on PPI network and the other on GO-based similarity network, and then they train a final joint DBN for prediction based on the outputs of the initial DBNs. In ~\citep{node_kernel2020}, the proposed approach Disjunctive Graph Integration (DiGI) merges gene co-expression network, pathways, functional links, phenotype similarity database, co-functional network and PPI network in a single network. Then, DiGI performs a novel node kernel on this single network, which is a decomposition kernel with two strategies, i.e., decomposition by the k-decomposition core and subsequently the clique decomposition. These techniques are used to extract features for each node to be fed into a regularized linear SVM for disease gene prediction.

There are also supervised machine learning endeavours, which work on tissue-specific networks \citep{Greene2015,Troyanskaya2018,Troyanskaya2012}. For instance, the proposed approach NetWAS \citep{Greene2015} combines functional interaction network of genes and hierarchically corrected tissue expression to obtain hierarchy-aware tissue-specific knowledge. Then, the tissue-specific knowledge and human data compendium are integrated through regularized Bayesian integration to form tissue-specific functional networks. Finally, using the constructed networks, tissue-specific disease analysis is performed to identify disease associations. Their platform Genome-Scale Integrated Analysis of Networks in Tissues (GIANT) interface \citep{Troyanskaya_Giant2018} further provides the tissue-specific maps and interactive visualizations.

As a final remark for supervised methods on networks, many of them have already employed the implicit semi-supervised learning assumption that samples close to each other tend to share the same labels. That is, the network structures provide intrinsic relationships between nodes (i.e., genes/proteins, diseases, etc.), which reveal valuable insights on related nodes and their tendency to associate with similar labels. The network structures are essentially label-free data that can aid supervised learning. Furthermore, additional multi-omics data (e.g., PPI, GO, gene expression, protein complexes, etc.) can be easily integrated into networks to further improve the performance of supervised learning.

\subsubsection*{PU learning methods}

Supervised models in disease gene prediction may suffer due to all provided evidence serves as positive samples and the negative samples remain unlabeled. Thus, positive-unlabeled (PU) methods arise to tackle this problem. They are usually based on weighting the positive samples and unlabeled samples to promote the prediction performance during the learning. ProDiGe \citep{Prodige} is a PU learning method for prioritization of candidate genes by enabling multiple data source integration such as PPI network and phenotype similarities through multitask learning strategy. In this way, scoring of the genes for prioritization is performed by considering disease-gene pairs instead of the individual genes. PUDI \citep{PUDI} partitions unlabeled set (i.e., negatives) into four sets namely, reliable negative set RN, likely positive
set LP, likely negative set LN and weak negative set WN based on features extracted from PPI, GO terms and protein domain. The weighted support vector machines are then used to build a multi-level classifier based on these sets and positive training set. CATAPULT \citep{Catapult} performs a biased SVM to classify gene-phenotype pairs on a combined heterogeneous gene-trait network including gene-phenotype, gene-gene interactions across multiple species. Features for SVM are derived from walks in the network. EPU~\citep{yang2014ensemble} is an ensemble-based PU learning method, which integrates data from multiple biological sources for training PU learning classifiers consists of weighted KNN, weighted naive Bayes and multi-level support vector machine. It starts with individual label propagation on gene expression network, PPI network and GO similarity network. Then, the obtained gene weights are combined and fed into the ensemble learner. PEGPUL \citep{jowkar2016perceptron} is an ensemble PU learning model to combine a multi-level SVM, a weighted KNN and a weighted decision tree for disease gene prediction. First, it extracts reliable negative genes by utilizing a co-training algorithm. Then, it constructs a similarity graph through metric learning by using PPI, protein domain and GO terms and performs a multi-rank-walk on the constructed graph to propagate labels prior to feeding the genes into the ensemble. A summary of supervised methods and PU learning methods is presented in Table~\ref{tab:handcrafted-feature}.

\subsection*{Graph representation learning methods}

Recall that we need to manually extract various features for the methods introduced in Table ~\ref{tab:handcrafted-feature}, which is tedious and requires domain knowledge. Recently, graph representation learning methods \cite{CuiEmbSurvey2017, cai2018comprehensive}, which can automatically learn the latent features/representations for the nodes, have acclaimed wide attentions in bioinformatics and biomedical applications \cite{SuEmbSurvey2018, yue2020graph}. In this section, we review graph representation learning methods designed for disease gene prediction. In particular, we divide them into four categories, namely, matrix factorization, graph embedding, graph neural network and hybrid methods as shown in Figure \ref{fig:Categories}.



\subsubsection*{Matrix factorization}
Matrix factorization (MF) techniques have been widely used for link prediction in bioinformatics, e.g., PPI prediction \cite{wang2013predicting}, drug-target prediction \cite{liu2016neighborhood, ezzat2016drug}, miRNA-disease association prediction \cite{chen2018predicting,zhang2020graph}, drug-pathway association prediction \cite{wang2020drug}, etc. Here, we introduce various MF methods developed for disease-gene association prediction. Basically, MF methods in gene-disease association prediction, such as inductive matrix completion (IMC)~\citep{IMC}, probability-based collaborative filtering (PCFM)~\citep{Zeng2017} and manifold learning \citep{Manifold2018}, aim to learn the latent factors for diseases and genes from the gene-disease association matrix. In addition, they can also specify the factorization process by including additional constraints into the objective function and thus MF techniques are useful for revealing important associations between diseases and genes.


 In \citep{IMC}, IMC constructs gene and disease features using different sources including human gene–disease associations, gene-expression from different tissue samples, functional interactions between genes, gene–phenotype associations of other species, disease similarities and disease-related textual data from OMIM database. Then, they form the gene–disease association matrix using these features and formulate an optimization problem to recover unknown low-rank matrix using observations from the constructed associations matrix. Their inductive approach is capable of making predictions for a query disease with no previously known gene associations. In \citep{Zeng2017}, PCFM exploits the gene–gene, gene–disease relationships, gene-disease linkages between orthologous genes and eight non-human species diseases and disease–disease similarity associations. They propose two probability-based collaborative filtering models, one with an average heterogeneous regularization and another with personal heterogeneous regularization using vector space similarity to predict gene-disease associations. The advantage of PCFM than the collaborative filtering is that PCFM enables probabilistic consideration in gene-disease associations instead of binary evaluation. Manifold learning \citep{Manifold2018} uses gene-disease association data for prediction and assumes that the geodetic distance between any associated gene-disease pairs are shorter than non-associated gene-disease pairs in a lower dimensional manifold. An optimization function is defined based on this assumption and singular value decomposition is employed to solve the optimization problem. Collage \citep{vzitnik2015gene} applies collective matrix factorization (CMF) to combine a wide range of 14 data sets including RPKM-normalized RNA-seq transcriptional profiles and phenotype ontology annotations. Then, it performs chaining on the learned latent matrices to obtain gene profiles for prioritizing bacterial response genes in Dictyostelium. Similarly, Medusa ~\citep{MEDUSA2016} also builds a collective matrix factorization model for data fusion of a large-scale collections of heterogeneous data and performs chaining on the learned latent matrices to establish connections between non-neighboring nodes in the fusion graph. It formulates the growing of the modules as a sub-modular optimization program. The proposed method is capable of both associating genes with diseases and detecting disease modules. In \citep{xi2018novel}, an unsupervised learning model based on matrix tri-factorization (tri-NMF) framework is proposed to detect disease causing genes from pan-cancer data. In particular, it exploits both the similarities of mutation profiles of different cancer types and gene interaction network. A method called GeneHound \cite{zakeri2018gene} adopts Bayesian probabilistic matrix factorization (BPMF) for disease gene prioritization problem. GeneHound uses gene-disease associations as partially observed data and a raw fusion is employed to integrate multiple genomic data sources including literature-based phenotypic and literature-based genomic information. Then, the Bayesian data fusion model jointly learns gene and disease latent factors and corresponding gene and disease-association matrices for disease-gene association prediction.

\subsubsection*{Graph embedding methods}

Graph embedding methods learn low-dimensional and continuous vector representations of nodes through a neural network. For example, SkipGram \citep{SkipGram} architecture is an extensively used architecture to construct associations between the node and its neighborhood. The neighborhood of the nodes is extracted through the random walks. Endeavours \citep{node2vec,DeepWalk,LINE}, such as DeepWalk \cite{DeepWalk} and node2vec \citep{node2vec}, generate node representations such that the nodes lying within the short random walk distance have similar embeddings. There are also several random walk-based embedding methods proposed for disease gene prediction as follows.

SmuDGE \citep{alshahrani2018semantic} combines disease–phenotype and gene–phenotype associations with interactions between genes to generate a corpus for SkipGram-based representation learning. Then, it builds an artificial neural network (ANN) to predict gene–disease associations. In \citep{xiong2019heterogeneous}, HeteWalk builds a weighted heterogeneous network by joining six public data sources including PPI, miRNA similarity network, and disease phenotype similarity network and then performs SkipGram based network embedding. It further applies meta-path selection to eliminate potential redundant and misleading information caused by the heterogeneous walks with multiple entities. HerGePred \citep{HerGePred2019} also performs SkipGram-based graph embedding on a heterogeneous network consisting of a PPI, disease-protein associations, protein-GO associations, and gene-disease associations. Eventually, HerGePred predicts novel disease-gene associations in two different manners. It can directly calculate the cosine similarity between the embedding vectors of the query disease and the proteins. It can also perform random walk on disease-gene network, which is reconstructed based on the calculated cosine similarities between embeddings.

\begin{table*}[t]
\centering
\caption{Graph representation learning methods for predicting disease-gene associations.}
\begin{tabular}{llp{11cm}}  \hline
Categories & Methods & Method descriptions \\\hline
\multirow{6}{*}{Matrix factorization} 
& IMC \citep{IMC} & Inductive matrix completion incorporating gene and disease features from multiple data sources.\\
& PCFM \citep{Zeng2017} & Probability-based collaborative filtering models with different regularization terms. \\
& Collage \citep{vzitnik2015gene} & Collective matrix factorization using 14 datasets including genes, GO terms, KEGG pathways, etc.\\
& Medusa \citep{MEDUSA2016} & Collective matrix factorization on a data fusion graph with 16 data matrices.\\
& Tri-NMF \citep{xi2018novel} & Matrix tri-factorization on mutation profile similarities for different cancer types and PPI network.\\
& GeneHound \cite{zakeri2018gene} &  Bayesian matrix factorization on OMIM associations with genomic and phenotypic data sources.\\
\hline
\multirow{3}{*}{Graph embedding}  
& SmuDGE \citep{alshahrani2018semantic} & RW-based embedding and an ANN on pair of disease and gene feature vectors. \\
& HeteWalk \citep{xiong2019heterogeneous}& RW-based embedding on a heterogeneous network with genes, miRNAs and diseases. \\
& HerGePred \citep{HerGePred2019} & RW-based embedding on a heterogeneous network with diseases, symptoms, genes and GO terms. \\\hline

\multirow{3}{*}{Graph neural networks}
& GCAS \citep{convoHANRD2018} & Graph convolution on heterogeneous association network for rare diseases (HANRD). \\
& PGCN ~\citep{li2019pgcn} & GCN to learn the embeddings for phenotypes and genes in a disease-gene heterogeneous network.\\
& VGAE ~\citep{singh2019towards} & VGAE to learn embeddings for diseases and genes in disease-gene networks for gene prioritization.\\
\hline

\multirow{6}{*}{Hybrid methods} 
& N2VKO \citep{Ata2018} & Combines node2vec embeddings and handcrafted features.\\
& N2A-SVM~\citep{Peng2019} & Combines node2vec embeddings with an autoencoder on a PPI network.\\
& N2Vmotif \cite{EmbeddingGraphlet} & High-order PPI structures combining node2vec embeddings and network motifs. \\
& GCN-MF \citep{GCNMF2019} & Combines GCN with matrix factorization using both gene similarities and disease similarities.\\
& HNEEM \citep{wang2019predicting} & An ensemble of 6 graph embedding methods in a disease-gene-chemical heterogeneous network.  \\
& DW-GCN  \citep{zhu2019predicting} &  Integrates graph embedding (DeepWalk) and graph convolutional network.\\
\hline\end{tabular}
	\label{tab:feature-learning}%
\end{table*}

\subsubsection*{Graph neural networks}

Graph neural network (GNN) is an advanced deep learning model for graph data \cite{wu2020comprehensive} and has been applied for various bioinformatics tasks \cite{SunCNNdrug2019, cai2020dual, long2020predicting}. Graph convolutional network (GCN), graph attention network (GAT) and graph auto-encoder (GAE) are representative GNN models. For example, GCN aims to learn node embeddings by implementing the convolution operation on a graph based on the properties of neighborhood nodes. Here, we will have a quick review on GNN models for disease gene prediction as follows.


In \citep{convoHANRD2018}, the authors introduce GCAS, which is an adaptation of graph convolutional network for disease gene prioritization. First, they construct a heterogeneous network consisting of ontological associations as well as curated associations including genes, diseases and pathways from multiple sources. They then perform direct spectral convolution to successively propagate the influence to the neighborhood and infer novel disease-gene associations. In ~\citep{Rao2020}, the authors introduce a tool PRIORI-T, which employs disease-gene, phenotype-phenotype and phenotype-disease correlation pairs extracted from a corpus of rare disease MEDLINE abstracts for gene prioritization. These correlations are computed based on Pearson correlation coefficient and used to construct initial correlation network (ICN). Then, GCAS~\citep{convoHANRD2018} is performed for gene prioritization on this network to obtain ranked gene list for each clinical case.  In \citep{li2019pgcn}, a graph convolutional network-based disease gene prioritization method called PGCN exploits a heterogeneous phenotype-gene network and the additional information for the nodes (e.g., disease ontology similarity, gene expression data, etc) for disease gene prioritization. In \citep{singh2019towards}, the authors introduce the variational graph auto-encoder (VGAE) for gene-disease association prediction in a heterogeneous disease-gene network.   

\subsubsection*{Hybrid methods}
Here, we denote the methods, which combine the representation learning methods with other techniques to derive features, as hybrid methods.


A number of methods combine graph embedding (e.g., node2vec) with other methods for feature engineering. In \citep{Ata2018}, a method called N2VKO integrates the node2vec embeddings extracted from PPI network with the biological annotations for disease-gene association prediction. In \cite{Peng2019}, the proposed method N2A-SVM employs node2vec embedding of the genes from a PPI network and then performs dimension reduction with auto-encoder to predict Parkinson’s disease genes. In \citep{EmbeddingGraphlet}, the authors propose to combine graphlet representations with node2vec embeddings for disease gene prediction.

\cite{GCNMF2019} builds a network-based framework for gene-disease association prediction, which exploits disease similarity and gene similarity graphs to construct two GCNs separately for diseases and genes. Then, GCNs are trained by using corresponding disease and gene features and optimized with label information with matrix factorization. In \citep{wang2019predicting}, the authors propose a heterogeneous network embedding method, HNEEM, for gene-disease association prediction by ensemble learning. The constructed heterogeneous network consists of gene-disease associations, gene-chemical associations ans disease-chemical associations. HNEEM first extracts graph embeddings with six embedding methods and then feeds these embeddings into a random forest classifier for disease gene prediction. DW-GCN \citep{zhu2019predicting} is proposed to combine DeepWalk and GCN for gene-disease association prediction on a heterogeneous disease-gene network. Final predictions are derived using the output of a GCN  decoder and the probability distribution derived from DeepWalk. Table~\ref{tab:feature-learning} summarizes various graph representation learning models for disease gene prediction.

\vspace{-0.2cm}

\section*{Empirical Comparison} \label{sec:emprical_eval}

In this section, we conduct experiments to evaluate the performance of various network-based methods for disease gene prediction.


\subsection*{Experimental Setup}

\subsubsection*{Datasets}
We collect two networks for proteins, namely, a PPI network from the IntAct database \citep{IntAct} and a protein functional similarity network based on GO terms \citep{Consortium2015}. In particular,  we first calculate the pairwise functional similarity between proteins based on G-SESAME\footnote{G-SESAME: http://bioinformatics.clemson.edu/G-SESAME/}~\citep{Wang2007}, and subsequently build a graph using the $K$ nearest neighbor (KNN) algorithm. We set $K=10$ in all of our experiments as its GO KNN graph can help node2vec to achieve the best performance for disease gene prediction. More details about GO KNN graph construction can be found in our supplementary materials. Overall, there are a total of 12,901 nodes (i.e., proteins) in both networks, with 96,845 edges in the PPI network and 107,508 edges in protein functional similarity network.

There are several publicly available databases for gene-disease associations as shown in Table~\ref{tab:databases}. In our experiments, we acquired the associations from the OMIM database. Given a specific disease/phenotype (e.g., Alzheimer's disease), we extract MIM IDs from OMIM Morbid Map and retrieve their corresponding protein IDs from the Uniprot~\citep{UniProt} conversion tool. Each protein node is subsequently assigned a binary label to represent whether it is a causative protein for this disease. In our experiments, we focus on seven diseases, namely, Alzheimer's disease (11), breast cancer (24), colorectal cancer (34), diabetes mellitus (37), obesity (19), lung cancer (15) and prostate cancer (17). Note that the number of positive proteins for each disease is included in the parentheses. The data and supplementary materials are available at https://github.com/sezinata/SurveyDGP.

\begin{table}[htbp]
  \centering
  \caption{Publicly available databases for gene-disease associations.}
    \begin{tabular}{lll}
    \hline
    Name  & URL   & Latest  Update \\
    \hline
    OMIM~\citep{OMIM2018}  &{https://omim.org} & July, 2020 \\
    DisGeNet~\citep{disgenet} & {https://www.disgenet.org} & June, 2020 \\
    MalaCards~\citep{MalaCards} & {https://www.malacards.org} & March, 2020 \\
    COSMIC~\citep{COSMIC} & {https://cancer.sanger.ac.uk} & April, 2020 \\
    PsyGeNET~\citep{PsyGeNET} & {http://www.psygenet.org} & January, 2018 \\
    CTD~\citep{CTD}  & {http://ctdbase.org} & July, 2020 \\
    \hline
    \end{tabular}%
  \label{tab:databases}%
\end{table}%

\subsubsection*{Selected methods for evaluation} \label{sec:benchmark}

We select a subset of methods from different categories for evaluation. For example, we select RWRH \citep{RWRH} from network diffusion methods and IMC \citep{IMC} from matrix factorization methods. For machine learning methods with hand-crafted features, we select Metagraph+ \citep{ata2017disease} from supervised methods, and Catapult \citep{Catapult} and ProDiGe \citep{Prodige} from PU learning methods. Furthermore, we select a graph embedding method node2vec \citep{node2vec} and a hybrid method N2VKO \citep{Ata2018}. 
    
As aforementioned, disease gene prediction is a typical node classification or link prediction task. Therefore, the state-of-the-art social network analysis methods for node classification or link prediction can be exploited for disease gene prediction. In our experiments, we also include HIN2Vec \citep{fu2017hin2vec}, HeGAN \citep{hu2019hegan}, MVE \citep{Qu2017}, mvn2vec \citep{shi2018mvn2vec}, DMNE  \citep{ni2018co} and MANE \citep{ata2020multiview} in our evaluation study. 

\begin{itemize}

\item \textbf{\emph{HIN2Vec}} \citep{fu2017hin2vec}: A heterogeneous network embedding approach, which samples heterogeneous paths called meta-paths and feeds them into a neural network. We combined PPI network (PPI view) and functional similarity network (GO view) to form a single heterogeneous graph, where edges from different views are assigned to a different type of relation.

\item \textbf{\emph{HeGAN}} \citep{hu2019hegan}: A heterogeneous network embedding approach, which utilizes the adversarial principle. We perform it on the constructed single network as described in \textit{HIN2Vec}. 

\item \textbf{MVE} \citep{Qu2017}: A state-of-the-art multi-view network embedding approach which maintain the collaboration between views by regularizing the Euclidean norm between view-specific embeddings and the final embeddings. The parameter $\eta$ is used to control the weight of regularization. Since we conduct our experiments on unsupervised models, we adopt the unsupervised version. 

\item \textbf{mvn2vec}~\citep{shi2018mvn2vec}: A state-of-the-art multi-view embedding approach. There are two proposed versions. Since they both have similar results we report only mvn2vec-r version which regularizes the Euclidean norm between view-specific embeddings, controlled by a hyperparameter $\gamma$. When $\gamma$ is set to zero, it becomes equivalent to node2vec performed on a single view. 

\item \textbf{\emph{DMNE}} \citep{ni2018co}:  A multi-view network embedding algorithm which is also capable of generating embeddings for many-to-many node mappings across views. Note that in our dataset only one-to-one mappings exist. We adopt their proximity disagreement formulation, due to its flexible assumption and better empirical performance. 

%
\item \textbf{\emph{MANE}} \citep{ata2020multiview}: A random-walk sampling-based multi view embedding algorithm, which unifies diversity, first-order collaboration and novel second-order collaboration principles in a framework. There are two hyperparameters $\alpha$ and $\beta$ to regularize the contribution of the principles.
\end{itemize}

The datasets that we utilize for the studied methods in this survey are as follows. node2vec and RWR are simply performed on the PPI network. In IMC, Catapult, ProDiGe and RWRH we use the PPI network, phenotype similarity network~\citep{mimminer} and the protein-phenotype associations. Metagraph+ and N2VKO leverage the PPI network and protein-keyword associations retrieved from Uniprot~\citep{UniProt}. HIN2Vec, HeGAN, MVE, mvn2vec-r, DMNE and MANE utilize GO and PPI networks described in Datasets section. For methods using protein-phenotype associations, we eliminate the test data associations to prevent data leakage.  


All methods are performed using the implementations provided by their respective authors unless stated otherwise, and we apply their suggested parameter settings for the hyperparameters. Table~\ref{tab:selectedMethods} shows the source code availability of the selected methods. Next, we briefly summarize the advantages and limitations of these selected methods. IMC, which is an MF method, basically aims to learn the latent factors of gene-disease association matrix. It is capable of predicting disease genes even for a disease that has no known associated genes. However, it generally does not provide global optimal solutions and has difficulty in convergence even for local optimal solutions \cite{MFtakestime2014}. Catapult and ProDiGe are useful for unbalanced data through biased SVMs, while they usually require tuning effort. RWRH is a random-walk based method. Although random-walk based methods are powerful in disease gene prediction, their performance might depend on the restart probability. Metagraph+ aims to identify candidate disease genes by capturing similarities through heterogeneous subgraphs incorporating both protein interactions and attributes. However, it is an arduous task to extract subgraph-level features (i.e., metagraphs). node2vec, N2VKO, mvn2vec-r, MVE and MANE are all subject to random walk-based sampling performance. However, mvn2vec-r, MVE and MANE might be able to provide more robust embeddings with the advantage of exploiting multi-view networks. HIN2Vec and HeGAN have the advantage of incorporating meta-paths. However, the former demands more training cost to capture the similarities between nodes and the latter requires a deliberate choice of pre-trained embeddings for initialization.

We adopt two well-known metrics for performance evaluation, namely, area under ROC curve (AUC) and area under precision-recall curve (AUPR). The performance of the models is evaluated through a stratified five-fold cross-validation. Given a specific disease, its causing genes are considered as positive samples and all the remaining genes as negatives. In particular, we randomly divide the positive and negative samples into five groups. For each round, we select one group of positive and negative samples as testing data and the other four groups as training data to calculate AUC/AUPR. Eventually we report the average AUC/AUPR over five rounds as the final AUC/AUPR score. For multi-view graph embedding models (i.e., MVE, mvn2vec-r, DMNE and MANE) we set walk length to 10, number of walks per node to 5, negative sampling size to 10, windows size to 3 and random walk parameters of node2vec $(p,q)$ to 1. All methods have $D=128$ as the dimension of the final embedding. Differently, for heterogeneous graph embedding model HIN2Vec we set walk length to 30 and number of walks per node to 10. For graph embedding methods (i.e., HIN2Vec, HeGAN, MVE, mvn2vec-r, DMNE and MANE), the learned final embeddings are fed into the logistic regression model. 

\begin{table}
      \caption{Selected methods for disease gene prediction and their source codes.}
    \begin{tabular}{p{1.5cm}p{6.3cm}}
          \hline
   Methods  & Source Code Availability \\\hline
    IMC   & https://bigdata.oden.utexas.edu/project/gene-disease\\
    Catapult  & http://www.marcottelab.org/index.php/Catapult \\ 
    ProDiGe  & http://cbio.ensmp.fr/prodige \\ 
    RWRH  & https://github.com/alberto-valdeolivas/RWR-MH \\ 
    Metagraph+ & https://github.com/sezinata/Metagraph \\ 
    node2vec & https://github.com/aditya-grover/node2vec \\
    N2VKO & https://github.com/sezinata/N2VKO \\  \hline

    HIN2Vec  & https://github.com/csiesheep/hin2vec \\
    HeGAN  & https://github.com/librahu/HeGAN \\
    MVE    & https://github.com/mnqu/MVE \\
    mvn2vec-r &  https://github.com/sezinata/mvn2vec-code  \\
     DMNE   & https://github.com/nijingchao/dmne \\
    MANE  &   https://github.com/sezinata/MANE\\  \hline
    \end{tabular}
\label{tab:selectedMethods}
\end{table}



\subsection*{Results and Discussion}
\label{subsection:ResultsandDiscussion}

\begin{table*}[htbp]
  \centering
  \caption{AUC performance comparison among the benchmark methods.}
\begin{tabular}{|c|l|cccccccc|}
\cline{1-10}\multicolumn{1}{|r|}{} & \multicolumn{1}{l|}{Disease} & \multicolumn{1}{c}{Alzheimer} & \multicolumn{1}{c}{Breast Cancer} & \multicolumn{1}{c}{Colon Cancer} & \multicolumn{1}{c}{Diabetes} & \multicolumn{1}{c}{Lung Cancer} & \multicolumn{1}{c}{Obesity} & \multicolumn{1}{c}{Prostate Cancer} & \multicolumn{1}{c|}{Avg} \\
\hline
\multirow{2}{*}{Homogeneous Network} & RWR   & 0.7179 & 0.7606 & 0.6241 & 0.5446 & 0.4842 & 0.5056 & 0.5047 & 0.5917 \\
      & node2vec & 0.7539 & 0.8997 & 0.8370 & 0.5103 & 0.6004 & 0.5500 & 0.5888 & 0.6772 \\
\hline
\multirow{8}{*}{Heterogeneous Network} & N2VKO & 0.8724 & 0.8201 & 0.8469 & 0.7477 & 0.7135 & 0.6427 & 0.6474 & 0.7558 \\
      & RWRH  & 0.8518 & 0.8857 & 0.9013 & 0.6138 & 0.7850 & 0.5544 & 0.5655 & 0.7368 \\
      & IMC   & 0.6688 & 0.7598 & 0.6618 & 0.6145 & 0.7555 & 0.6221 & \textbf{0.8185} & 0.7001 \\
      & Catapult & 0.8241 & 0.8258 & 0.8748 & 0.6211 & \textbf{0.8431} & 0.6144 & 0.5762 & 0.7399 \\
      & ProDiGe & 0.9041 & 0.8997 & 0.8875 & 0.6525 & 0.7943 & \textbf{0.7856} & 0.5107 & 0.7763 \\
      & Metagraph+ & 0.8978 & 0.7725 & 0.8981 & 0.7335 & 0.6709 & 0.5999 & 0.6702 & 0.7490 \\
      & HIN2Vec &0.7858  &0.8775 & 0.9225 &0.6995 & 0.7178& 0.6850&0.6272  &0.7593  \\
      & HeGAN & 0.8423 & 0.9199 & 0.9036 & 0.6677 & 0.7224 & 0.5765 & 0.5131 & 0.7351 \\
\hline
\multirow{4}{*}{Multi-view Network} & MVE   & 0.6543 & 0.8330 & 0.8520 & 0.6305 & 0.5722 & 0.4078 & 0.5339 & 0.6405 \\
      & mvn2vec-r & 0.8756 & 0.9002 & 0.9187 & 0.6489 & 0.6482 & 0.6692 & 0.5088 & 0.7385 \\
      & DMNE  & 0.9357 & 0.7996 & 0.8496 & \textbf{0.7526} & 0.7152 & 0.7256 & 0.4727 & 0.7501 \\
      & MANE  & \textbf{0.9660} & \textbf{0.9276} & \textbf{0.9244} & 0.7157 & 0.6951 & 0.7339 & 0.6069 & \textbf{0.7956} \\
\hline
\end{tabular}%
  \label{tab:AUC}%
\end{table*}%

\begin{table*}[htbp]
  \centering
  \caption{AUPR performance comparison among the benchmark methods.}
    \begin{tabular}{|c|l|cccccccc|}
\cline{1-10}    \multicolumn{1}{|r|}{} & Disease & Alzheimer & Breast Cancer & Colon Cancer & Diabetes & Lung Cancer & Obesity & Prostate Cancer & Avg \\
 \hline
    \multirow{2}[1]{*}{Homogeneous Network} & RWR   & 0.0063 & 0.0321 & 0.0140 & 0.0154 & 0.0049 & 0.0530 & 0.0123 & 0.0197 \\
          & node2vec & 0.0411 & 0.0682 & 0.0749 & 0.0048 & 0.0046 & 0.0046 & 0.0060 & 0.0292 \\
    \hline
    \multirow{8}[2]{*}{Heterogeneous Network} & N2VKO & 0.1592 & 0.0776 & 0.1185 & 0.0392 & 0.1083 & 0.0618 & 0.0357 & 0.0857 \\
          & RWRH  & 0.1760 & 0.1057 & 0.1411 & 0.0085 & \textbf{0.1203} & \textbf{0.1322} & 0.0042 & 0.0983 \\
          & IMC   & 0.0058 & 0.0094 & 0.0063 & 0.0056 & 0.0103 & 0.0073 & 0.0089 & 0.0076 \\
          & Catapult & \textbf{0.3537} & 0.0718 & 0.0845 & 0.0056 & 0.0761 & 0.0109 & 0.0076 & 0.0872 \\
          & ProDiGe & 0.3420 & 0.0732 & 0.0481 & 0.0154 & 0.1114 & 0.0384 & \textbf{0.0532} & 0.0974 \\
          & Metagraph+ & 0.3018 & 0.1308 & 0.1053 & 0.0159 & 0.0383 & 0.0679 & 0.0032 & 0.0947 \\
          & HIN2Vec & 0.0165 & 0.0987 & 0.0972 & 0.0135 & 0.0517 & 0.0042 & 0.0178 & 0.0428 \\
          & HeGAN & 0.1442 & 0.1167 & 0.1091 & 0.0129 & 0.0101 & 0.0047 & 0.0028 & 0.0572 \\
    \hline
    \multirow{4}[2]{*}{Multi-view Network} & MVE   & 0.0161 & 0.0622 & 0.0757 & 0.0283 & 0.0150 & 0.0016 & 0.0405 & 0.0342 \\
          & mvn2vec-r & 0.0275 & 0.0868 & \textbf{0.1570} & 0.0072 & 0.0072 & 0.0794 & 0.0023 & 0.0525 \\
          & DMNE  & 0.0603 & 0.0205 & 0.0252 & 0.0123 & 0.0503 & 0.0877 & 0.0018 & 0.0369 \\
          & MANE  & 0.2277 & \textbf{0.1889} & 0.1252 & \textbf{0.0435} & 0.0850 & 0.0386 & 0.0084 & \textbf{0.1025} \\
    \hline
    \end{tabular}%
    
  \label{tab:AUPR}%
\end{table*}%


We demonstrate the performance comparison of 14 state-of-the-art methods in Table~\ref{tab:AUC} and Table~\ref{tab:AUPR} in terms of AUC and AUPR, respectively. In particular, we group these methods into three categories based on their most proper input data to provide more insight about the approaches, namely homogeneous graph, heterogeneous graph and multi-view graph. In our supplementary materials, we also performed over-sampling with SMOTE \citep{chawla2002smote} for the network embedding methods. The oversampling results as well as the comparison results in terms of ranking-aware metrics are provided in our supplementary materials. Overall, the AUC and AUPR scores in Table~\ref{tab:AUC} and Table~\ref{tab:AUPR} are generally correlated, i.e., a method performing better in AUC tend to perform better in AUPR as well, with a Pearson correlation coefficient of about 0.7. In addition, AUPR results are much lower compared to AUC due to the skewness of the datasets, and this is common for disease gene prediction in several studies \cite{xi2018novel, GCNMF2019, wang2019predicting}. Based on the results in Table~\ref{tab:AUC} and Table~\ref{tab:AUPR}, we can have the following observations. 

Firstly, we can observe that RWR and node2vec working on a homogeneous PPI network achieve low performance in terms of both AUC and AUPR as expected. Therefore, we are motivated to integrate different data sources into a heterogeneous network or multi-view network to improve the prediction performance.


Secondly, various methods working on a heterogeneous network perform well for disease gene prediction. ProDiGe and Catapult, as PU learning methods, are very promising to address the imbalanced data issue for disease gene prediction and both incorporate the phenotype similarity network and phenotype-gene associations. In particular, ProDiGe achieves the second best average AUC of 0.7763 and the third best average AUPR of 0.0974 among the 14 selected methods. In addition, RWRH also achieves a stable performance and thus it is a good alternative to the network embedding methods. However, RWRH's performance depends on the tuning process, e.g., the setting of the restart probability as demonstrated in the supplementary materials. N2VKO and Metagraph+ first work on feature extraction/selection in a PPI and keyword network, and then undergo an oversampling procedure. Both of the methods achieve a decent performance. Note that the random walk parameters of node2vec $(p,q)$ in N2VKO are not tuned to maintain consistency with random walk sampling based embedding methods. Heterogeneous graph embedding methods including HIN2Vec and HeGAN hinge on different principles and perform very well. HIN2Vec utilizes random walks and negative sampling with a neural network model, while HeGAN utilizes adversarial learning principle in which a discriminator and a generator compete with each other as in a mini-max game. In particular, HeGAN's performance highly depends on the used pre-trained embeddings as inputs and thus careful initialization is critical in this model.  


Lastly, multi-view methods are also robust and useful tools for disease gene prediction. mvn2vec-r and MANE are two random walk-based embedding methods employing Skip-gram for learning the final embeddings, while DMNE utilizes a deep autoencoder in place of the Skip-gram architecture and also employs RW-based sampling to feed into the autoencoder. They can achieve relatively good performance in terms of AUC. In particular, MANE introduces a novel second-order collaboration and combines it with the previously studied principles in a unified framework. Therefore, it has achieved the best performance in average and outperforms ProDiGe and HIN2Vec by 2.5\% and 4.8\% in AUC, respectively. Meanwhile, MVE is an attention-based supervised algorithm and we employ its unsupervised version for a fair comparison in our experiments. It is thus reasonable to achieve a relatively low performance for MVE.

\section*{FUTURE PERSPECTIVES}
\label{sec:Future_work}

In this section, we present possible future directions that may address current challenging issues for more accurately predicting disease genes.

\subsection*{Learning with Limited Labeled Data}

Learning with limited labelled data has been a challenging task in disease gene prediction. Existing efforts to overcome this difficulty include PU learning and oversampling techniques. For example, PU learning methods ~\citep{Prodige,Catapult,PUDI} select likely-positive samples from the unlabeled data to tackle the problem. Oversampling of minority class samples (e.g., SMOTE \cite{chawla2002smote}) is also a common strategy to address this challenge. However, both strategies might need a tremendous tuning effort for difficult scenarios to attain a satisfactory performance. Developing more efficient and accurate models leveraging these strategies could be a promising direction.

Recently, generative adversarial networks (GAN) have been successfully applied to augment the data for various tasks, e.g., image classification \cite{gurumurthy2017deligan, frid2018gan}, speech recognition \cite{saito2017statistical}, etc. It is thus worth investigating GAN-based techniques for disease gene prediction with limited labels in the graph data. In addition, researchers also employ GAN to boost the PU learning \cite{hou2018generative} and oversampling processes \cite{mullick2019generative}. Therefore, it would be very interesting to see the efforts that combine GAN with PU learning or oversampling for disease gene prediction in the future.

\subsection*{Attention Mechanisms and Data Integration}

Graph Attention Networks (GAT) \cite{Velickovic2018}, as extension of graph convolutional networks, assign different weights to different neighbours with masked self-attention layers. In particular, this self-attention operation enables the model to focus on more important neighbours. Due to the attention mechanism, GAT have been applied to generate accurate graph embeddings for various tasks in recommendation systems \cite{wu2019dual, wang2019kgat} and bioinformatics \cite{attentiongraphandseqintoCNN2018}. Therefore, it is also promising to develop node-level attention mechanisms for diseases and genes in different types of graph inputs for network-based disease gene prediction.

Moreover, it is common to integrate different graph data sources through multi-view techniques \cite{li2018review}. In these multi-view techniques, it is important to reveal the contributions of each network to the final prediction performance, so that it would be possible to prioritize the networks based on their significance. Graph-level attention mechanisms \cite{ata2020multiview} can be useful in multi-view graphs for this purpose. In addition, hierarchical attention mechanisms \cite{yang2016hierarchical} can also be applied for disease gene prediction by combining both node-level and graph-level attentions.  

\subsection*{Sampling Strategies for Multi-View Inference}

The study N2VKO \citep{Ata2018} demonstrates that node2vec embeddings achieve an inferior prediction performance on some diseases such as the obesity and prostate cancer. The reason is that the disease-associated proteins are scattered in the network with a greater hop-distance between them, which limits the prediction power of a RW-based sampling method (e.g., node2vec). In these cases, it is good to develop models that can utilize other structures in the network. Specifically, this problem can be tackled by sampling strategies which may not consider only the local neighborhood of a node in a network. For example, we can adopt a collaborative sampling strategy, which can consider all the views of multi-view networks or utilize the attributed nodes during the sampling procedure. It would thus minimize the effect of local neighborhoods and help to increase the prediction performance. 

\subsection*{Single-cell Data}

Recently, single-cell RNA-seq (scRNA-seq) techniques become popular with the advancements in next-generation sequencing. Compared to traditional bulk RNA-seq analysis, they enable cell-level sequencing to capture cell-to-cell heterogeneity. In particular, several methods have been developed to infer gene relationships \cite{yuan2019deep} and gene regulatory networks \cite{iacono2019single} from single-cell gene expression data. Disease gene prediction can thus benefit from such inferred gene relationships and gene regulatory networks.

In addition, precise identification of cell states and types is crucial to understand the disease related mechanisms so that the scientists can detect correlated expression levels of genes across a homogeneous population of cells \citep{Trapnell01102015}. Community detection algorithms such as louvain \citep{louvain2008} and its similar version leiden \citep{Traag2019} are very popular tools to identify cell clusters. Furthermore, it is possible to model single-cell gene expression data as graphs and we can thus employ graph embedding methods for cell type identification through graph clustering \citep{SCRL2017}. 

\subsection*{Explainable Machine Learning Models}

Decision trees, linear regressions and logistic regressions are commonly used explainable machine learning models. The rise of the neural networks and deep learning in many areas such as video, speech and text processing comes with the need of explanation for the ``black box'' nature of these models. As we know, it is very important for medical disease-related predictions to provide information on why the model performs a certain prediction. However, the graph representation learning methods covered in this review are also lack of interpretability. We are thus highly motivated to develop explainable machine learning models for disease gene prediction.

Recently, knowledge graphs (KG) have been integrated with user-item graphs for accurate and explainable recommendation \cite{wang2019kgat, wang2019knowledge}. A recent work \cite{ijcai2020-380} adopts KG and graph neural networks for explainable drug-drug interaction (DDI) prediction. Similarly, we can also construct knowledge graphs for diseases and genes using various side information. For example, we can obtain the medical KG with diseases, drugs and symptoms, and gene ontology KG with genes and their functional annotations. By integrating advanced graph neural network techniques (e.g., GAT) and knowledge graphs, we expect to achieve accurate and explainable predictions for novel disease-gene associations.

\section*{Conclusion} \label{sec:Conclusion}

Discovering disease causing genes and analyzing their roles in the disease are not only critical for understanding disease formation mechanism, but also extremely important for designing appropriate drugs for corresponding clinical therapies. 
Linkage analysis and genome-wide association studies (GWAS) form the basis of disease gene  prediction. However, they generate a large number of false positives in their statistical analysis of biomarkers. 
Computational approaches are efficient and complementary tools to help biologists  filter out noisy false positives and provide a list of genes which are worth for further clinical study. In this survey, we focus on network-based research, leveraging various networks in their problem formulation for disease gene prediction. We provide an organized, up-to-date overview of state-of-the-art network-based approaches. We also perform an empirical comparison study on different computational methods based on different graph inputs. 

Generally, the methods in both heterogeneous network and multi-view network perform very well for disease gene prediction. In particular, multi-view methods with Skip-gram architecture can serve as robust and useful tools, not only for disease gene prediction but also for visualization and clustering purposes. Its low computational cost with a minimal tuning necessity and high prediction performance, make it a good alternative to the other learning methods. These techniques could be further investigated by considering  attention mechanisms 
or different network sampling strategies instead of random walks. Moreover, for network-based disease gene prediction analysis, constructing reliable networks is critical, so future research need to focus on predicting novel interactions and removing noisy interactions. The ultimate success of the disease gene prediction will depend on the \textit{parallel improvements} both in the experimental techniques by biologists and clinicians to provide rich and reliable biological
data sets, and in the advanced computational techniques by computer scientists to provide efficient and robust ways to discover novel knowledge and insights from the biological data.\\

\section*{Biographical Note}
\textbf{Sezin Kircali Ata} just obtained her PhD degree from School of Computer Science and Engineering Nanyang Technological University (NTU), Singapore and currently is a research fellow at KK Women's and Children's Hospital, Singapore. Her research interests include machine learning, graph mining and bioinformatics. 
\textbf{Min Wu} is currently a senior scientist at the Institute for Infocomm Research (I2R), A*STAR, Singapore.  His research interests include machine learning, data mining and bioinformatics. 
\textbf{Yuan Fang} is currently an Assistant Professor at the School of Information Systems, Singapore Management University, Singapore. His research focuses on graph-based machine learning, Web and social media mining and recommendation systems. 
\textbf{Le Ou-Yang} is an assistant professor in the College of Electronics and Information Engineering, Shenzhen University, Shenzhen 518060, China. His research interest includes bioinformatics and machine learning.
\textbf{Chee-Keong Kwoh} is currently an associate professor at the School of Computer Science and Engineering, NTU, Singapore. His research interests include data mining, soft computing and graph-based inference, bioinformatics and biomedical engineering.
\textbf{Xiao-Li Li} is currently a department head and principal scientist at I2R, A*STAR, Singapore. His research interests include data mining, machine learning, AI, and bioinformatics. \\

\framebox{

\parbox[t][5.5cm]{0.85\columnwidth}{
\color{black}
\bf{Key Points}
\begin{itemize}
\item Uncovering disease-causing genes is a fundamental objective of human genetics, while computational prediction of disease-genes provides a low-cost alternative. 

\item We classified and reviewed state-of-the-art network-based approaches for disease gene prediction with different types of graph inputs.

\item We empirically evaluated various selected methods, including some advanced methods for social network analysis, for disease gene prediction.

\item We also discussed possible  future  directions  that  may  address current challenging issues for more accurately predicting disease genes.

\end{itemize}
}
}

\bibliographystyle{unsrt}
\bibliography{ref} 

\end{document}